\documentstyle[prl,epsf,epsfig,amssymb,amsbsy,amstext,amsfonts,aps]{revtex}
\begin{document}
\twocolumn
\draft{}
\bibliographystyle{try}

\topmargin 0.1cm


\wideabs{

\def\slac{$^{1}$}
\def\jlab{$^2$}
\def\nordita{$^{3}$}
\def\sphn{$^4$}

\title{Compton scattering, vector meson  photoproduction and the partonic structure of the nucleon}

\author{F. Cano and J.M. Laget 
}


\address{
CEA Saclay, DAPNIA-SPhN, F91191 Gif-sur-Yvette Cedex, France
}
\date{\today}

\maketitle

\begin{abstract}
At moderate and large momentum transfer, high energy vector meson
photoproduction and Compton scattering can be consistently explained
by mechanisms based on interactions and exchanges of constituent
quarks and gluons. This opens an original window on the partonic
structure of hadronic matter by emphasizing the role of quark
correlations, quark exchange as implemented through saturating Regge
trajectories and non--perturbative effects in the gluon propagator,
which provides us with a bridge with lattice calculations. Compton
scattering is directly related to vector meson photoproduction and
enlarges the data set.
 
\end{abstract}

}

\narrowtext

Since a photon has the same quantum numbers as vector mesons, high energy photoproduction or electroproduction of Vector Mesons allow to prepare a beam of quark-antiquark pairs of a given flavor. It is selected by the kind of vector meson which is detected: light quarks for $\rho$ or $\omega$, strange quarks for $\phi$ and charmed quarks for $J/\psi$ mesons. The life time of the fluctuation $2\nu /(Q^2+M_V^2)$ is given by the uncertainty principle and increases with the beam energy $\nu$.   Therefore, the interaction with matter of a beam of energetic photons is similar to the interaction of hadrons~\cite{La01,La00}.

\begin{figure}[h]
\centerline{\epsfig{file=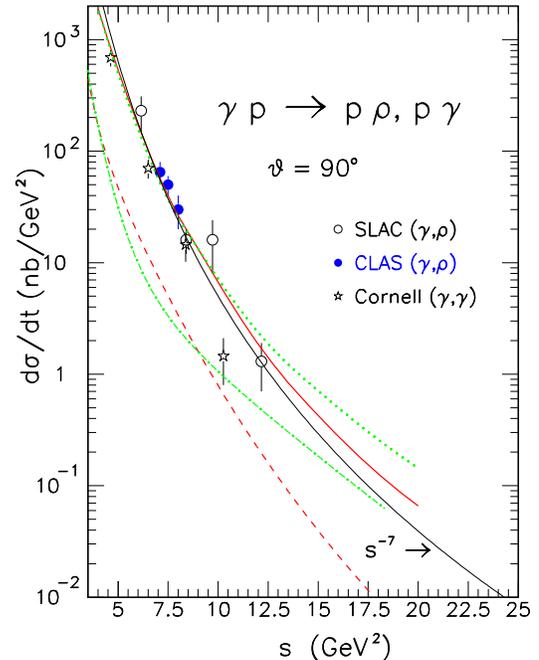,height=3.5in}}
\caption{The $\theta =90^{\circ}$ cross section of the $p(\gamma,
\rho)p$ reaction. Full (dashed) curve corresponds to the results of
the full model with a lattice--based (gaussian) propagator. Bottom
curves represent the contribution of two--gluon exchange only with
lattice (dashed line) or gaussian propagator (dash--dotted line). 
Thin solid line show the $s^{-7}$ behavior. Data for Compton
scattering has been renormalized by $f_V^2/4\pi \alpha_{em}$.} 
\label{rho_90}
\end{figure}

For the same reason, the energetic photon which is emitted in Compton scattering couples to the nucleon as a vector meson. Consequently, the Compton scattering cross section can be deduced by multiplying the $\rho$ meson photoproduction cross section, which dominates over the $\omega$ and $\phi$ ones,  by $4\pi \alpha_{em}/f_V^2$, where $f_V$ is the radiative decay constant of the vector meson. This is supported by the sparse available experimental data. Fig.~\ref{rho_90} compares the  renormalized cross section of  the $p(\gamma, \gamma)p$ reaction to the cross section of the $p(\gamma, \rho)p$ reaction, at $\theta =90^{\circ}$. 

Their steep decrease with $s$ (the squared total energy in the
c.m. frame) is consistent with the asymptotic behavior ($s^{-6}$ for
Compton scattering, $s^{-7}$ for rho photoproduction) inferred from
hard scattering on current quarks. However, a fully perturbative
calculation misses the Compton scattering cross section by about one
order of magnitude, unless unreasonable nucleon wave functions are
used (see Ref.~\cite{Va97}, for instance). More recently, it has been
realized that soft processes, which can be related to overlap integrals of non perturbative
parts of the nucleon wave function~\cite{Di99,Ra98,Hu00}, dominate the cross sections in this energy range.

In this note, we propose an alternative description of vector meson photoproduction and Compton scattering. It is based on effective partonic degrees of freedom, which can be tested against lattice gauge calculations: dressed gluons or constituent quarks propagators, parton correlations in the nucleon wave functions. The analysis of different channels allows one to assess the importance of the various players in the game. The expression of the various amplitudes may be found in Ref.~\cite{La00}.

\begin{figure}[h]
\centerline{\epsfig{file=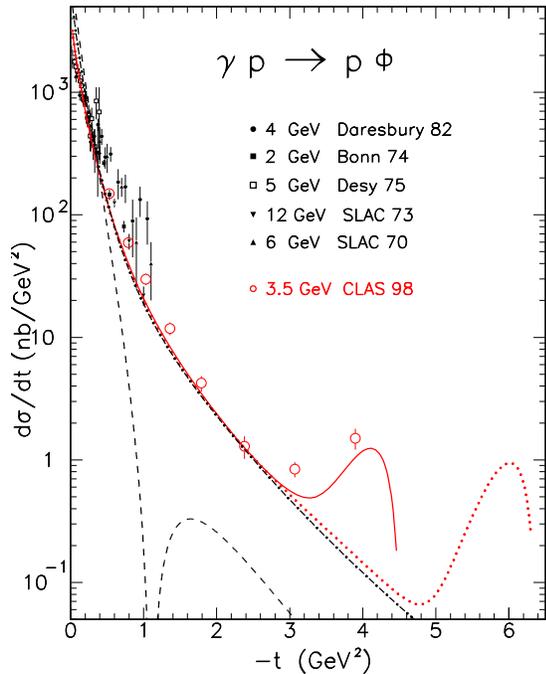,height=3.5in}}
\caption{The differential cross section of the $p(\gamma, \phi)p$, in
the JLab energy range: $E_\gamma=3.5$ GeV (solid line) and
$E_\gamma=4.5$ GeV (dotted line). Dash-dotted (dashed) line is the
contribution of the two-gluon exchange only with (without) quark
correlations.}
\label{phi_kroll} 
\end{figure}

$\phi$ meson photoproduction allows one to prepare a $s\bar s$ pair of strange quarks and study its interaction with hadronic matter~\cite{La00}. At low momentum transfer $t$ (small angle), its diffractive scattering is mediated by the exchange of the Pomeron. At high momentum transfer (large angle), the impact parameter is small and comparable to the gluon correlation length (the distance over which a gluon propagates before hadronizing): the Pomeron is resolved into its simplest component, two gluons which may couple to each of the quarks in the emitted vector meson or in the proton target. Configurations in the proton with small transverse size are selected and therefore each gluon can couple to a different quark. While these contributions are small at low $-t$, as argued in~\cite{LANDSHOFF87}, they become essentiel to get good agreement with data at high $-t$, as ilustrated in Fig.~\ref{phi_kroll}. Indeed, when
they are neglected the theoretical calculation show a node in the
differential cross section which is ruled out by experiments. In that region the cross section is sensitive to quark correlations in the proton. 

The coupling of two gluons to two different quarks in the nucleon 
is controled by the correlation function $G_2$ which, in the eikonal approach, reads: 

\begin{eqnarray}
G_2(\vec{k}_a, \vec{k}_b) & = & \int \Pi_i d x_i d \vec{r}_i \delta(\sum_i
x_i 
\vec{r}_i) \delta(\sum_i x_i -1) \nonumber \\
& & |\Psi(x_i,\vec{r}_i)|^2 
e^{i (\vec{k}_a \cdot \vec{r}_j + \vec{k}_b \cdot \vec{r}_k) } \; ,
\end{eqnarray}

\noindent where $\vec{k}_a$, $\vec{k}_b$ are the transverse momenta
carried by each gluon, $(\vec{k}_a + \vec{k}_b)^2 = -t$, and $\Psi$ is
the nucleon wave function; its arguments are $x_i$ the fraction of
longitudinal momentum carried by each quark, and $\vec{r}_i$ its
transverse coordinate. It is customary to approximate $G_2$ by the
isoscalar Dirac form factor $F_1$ with a shifted argument
\cite{CUDELL94}:

\begin{equation}
G_2(\vec{k}_a,\vec{k}_b) = F_1 (\vec{k}_a^2+ \vec{k}_b^2 - \vec{k}_a
\cdot \vec{k}_b ). 
\label{G2approx}
\end{equation}
  
This approximation has also been used in the previous
analysis~\cite{La00} of recent JLab~\cite{An00} and DESY~\cite{Br00}
data but it is strictly valid if the longitudinal
momentum of the proton is equally shared by the quarks, i.e. $x_i=1/3$
\cite{CUDELL94}. In order to
fully account for the longitudinal motion of quarks, we have evaluated
$G_2$ by using the functional form of the nucleon wave function
proposed by Bolz {\it et al.} \cite{BOLZ96}:

\begin{equation}
\Psi(x_i, \vec{r}_i) = N_\Psi \phi_{{\mbox  \small AS}} 
 \exp[-\frac{1}{4 a_N^2} \sum_{i} x_i \vec{r}_i\,^2]
\label{wavefunction}
\end{equation}

\noindent where $N_{\Psi}$ is a normalization constant, $\phi_{{\mbox
\small AS}}=120 x_1 x_2 x_3 $ is the asymptotic distribution amplitude
for the nucleon. This particular dependence on $x_i$ and $\vec{r}_i$
fulfills some fundamental properties of QCD \cite{ZHITNITSKY98} based on
duality and dispersion relations. It has been successfully used
to evaluate soft contributions to a number of processes \cite{Di99},
such as nucleon form factors, parton distributions and Compton Scattering
amplitudes. The only free parameter in our case, $a_N$, has been fixed to reproduce the isoscalar Dirac form factor up to a momemtum transfer $\lesssim$ 6 GeV$^2$, which gives   $a_N=1.3$ 
GeV$^{-1}$ and an averaged transverse momentum $\langle k_{i}^2
\rangle^{1/2}= 0.314$ GeV. As a general feature~\cite{CANO01}, the two-quark correlation $G_2$ evaluated with the wave function (\ref{wavefunction}) falls off at large momentum more rapidly than the one obtained with the ansatz~(\ref{G2approx}).
 
	Concerning the gluon propagator, there is now a firm evidence
from lattice calculations that gluons behave as  massive particles at
low momenta and non-perturbative effects may show up at momenta as
large as 4 GeV$^2$ \cite{WILLIAMS01}. Since the typical momentum
flowing through gluons are of the order $-t/4 \lesssim 1-3$ GeV$^2$
for our region of interest, non-perturbative effects in the gluon
propagators cannot be neglected and the use of a perturbative one is
questionable. 

In previous works~\cite{La00,LAGET95} we used the gaussian shape~\cite{LANDSHOFF87}:

\begin{equation}
\alpha_n D(l^2) = \frac{3 \beta_0}{\sqrt{2 \pi} \lambda_0} 
\exp [-l^2/\lambda_0^2]\; ,
\label{gaussian}
\end{equation}  

\noindent where the parameter $\beta_0=2$ GeV$^{-1}$ is the quark-pomeron coupling and $\lambda_0^2=2.7$ GeV$^2$ is fixed from phenomenological grounds  \cite{LAGET95}. This gives a correlation length for the gluons of 0.12 fm.  
$\alpha_n$ is interpreted as the (frozen) strong coupling constant  in the non--perturbative domain where one operates.  

Cornwall \cite{CORNWALL82} derived an infrared safe gluon propagator where
non-perturbative effects are taken into account through a dynamically
generated gluon mass

\begin{equation}
m^2(l^2) = m_g^2 \big[\log(\frac{4 m_g^2}{\Lambda^2})/\log(\frac{l^2
+4 m_g^2}{\Lambda^2})\big]^{12/11} \; .
\label{gluonmass}
\end{equation}
  
\noindent It vanishes at large momentum transfer, rendering to the
propagator its correct asymptotic perturbative behavior, but has a finite value at the origin $m_g$,  regularizing the amplitudes. Recent
parameterizations of lattice data \cite{LEINWEBER99} have also made use
of these  functional form, with a value of
$m_g=0.452$ GeV($\Lambda=0.508$ GeV), compatible with the estimate of
Cornwall $m_g=0.5\pm 0.2$ GeV. In this note we use
 a more accurate parameterization (model A in
ref.~\cite{LEINWEBER99}) of lattice data: 

\begin{equation}
D(l^2) = Z \left[ \frac{A M^{2 \alpha}}{(l^2)^{(1+\alpha)} +
(M^2)^{1+\alpha}} + \frac{L(l^2,M)}{l^2 + M^2}\right]   
\label{modela}
\end{equation}

The particular values for the parameters of the expression above can
be found in \cite{LEINWEBER99}.
The corresponding correlation length is 0.22 fm, close to the one given by the Cornwall-based parameterization, 0.24 fm. This value is a factor 2 larger than the correlation length provided by the gaussian propagator. Furthermore, the shape of the lattice-based propagator is very different from the gaussian one. It moves the node in the uncorrelated part of the cross-section in Fig.~\ref{phi_kroll} from $-t= 2.3$~\cite{La00} to $-t= 1.2$, and leads to a steeper dependency at large $-t$.

In the JLab energy range, $u$-channel nucleon exchange ``pollutes''
the highest $t$ bin: here the $\phi NN$ coupling constant $g_{\phi
NN}= 3$  is the same as in the analysis of the nucleon electromagnetic
form factors~\cite{Ja89}. More details may be found in
ref.~\cite{La00}. At JLab, the experiment has been repeated at higher energy (around E$_{\gamma}$ = 4.5~GeV). The $u$-channel backward peak is moved at higher values of $-t$ (dotted curve), leaving more room to reveal and check the two gluon exchange contribution. The preliminary results (not shown) confirm these predictions, showing the adequacy of the correct treatment of quark correlations, as well as of the lattice QCD calculations of gluon propagators. 

Such a finding is important as it tells us that, in the intermediate range of momentum  transfer (let's say $1\leq -t\leq 10$ GeV~$^2$), large angle exclusive meson production can be understood in a perturbative way at the level of effective parton degrees of freedom: dressed quark and gluon propagators, constituent quark wave functions of the nucleon and of the meson, etc\ldots.
At low momentum transfer (up to $-t\approx 2$~GeV$^2$), the cross section is driven by their integral properties: any nucleon wave function which reproduces the nucleon form factor leads to similar results; the differences which may be due to gluon propagators can be reabsorbed in a reasonable value of $\alpha_n$ (for the lattice based propagator we have used $\alpha_n=0.28$). At higher momentum transfer, the cross section becomes more sensitive to the details of the wave function (giving access to quark correlations) and the shape of the gluon propagator.  The large momentum cross section is reduced when either a more realistic wave function (than used in ref.~\cite{La00}), or a lattice-based gluon propagator (instead of a gaussian gluon propagator as in ref.~\cite{La00}), is used. 

\begin{figure}[h]
\centerline{\epsfig{file=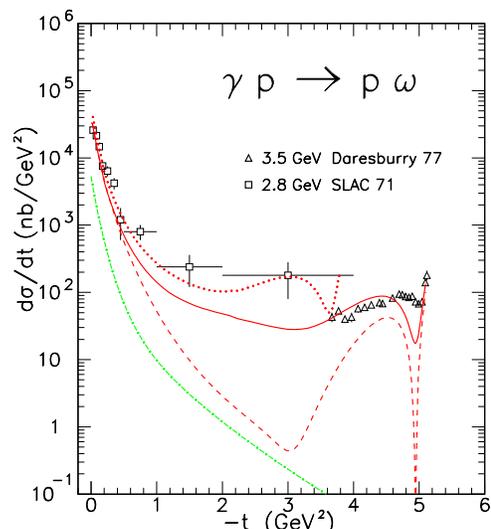,height=2.75in}}
\caption{The differential cross section of the $p(\gamma, \omega )p$
reaction, at $E_\gamma=2.8$ GeV (dotted line) and $E_\gamma=3.5$ GeV 
(solid line). Dashed line represents the later case when linear Regge
trajectories are used instead of saturating ones. Dash-dotted line is
the contribution of the two-gluon exchange only.}
\label{omega_kroll}
\end{figure}

In contrast to the $\phi$ meson sector, quark interchanges are not forbidden in the $\rho$ and $\omega$ meson photoproduction sector. 
Fig.~\ref{omega_kroll} shows the sparse data in the $\omega$ production channel. Pion exchange dominates the cross section and reproduces the forward angle cross section, while the exchange in the $u$-channel of the nucleon Regge trajectory reproduces the backward angle data. Note that the dip at hight $-t$ is due to the zero in the nucleon non degenerated Regge trajectory~\cite{Gui97}:  as in the $\phi$ channel, this is the only baryon trajectory which can be exchanged.  At intermediate angles, the two gluon exchange contribution badly misses (by an order of magnitude or more) the experimental cross section. It is driven by the same expression~\cite{La00} as in the $\phi$ channel, where only the relevant mass and radiative decay width have been changed.

The agreement is restored when one uses the pion Regge saturating trajectory ($\alpha_{\pi}^{sat}(t) \rightarrow -1$ when $t \rightarrow -\infty$) which reproduces  the cross section of the $\gamma p\rightarrow \pi^+ n$ reaction around $90^{\circ}$~\cite{Gui97}. 
It leads to the asymptotic quark counting rules, which model-independently determine the energy dependency of the cross section at large $-t$. In a world where  the mesons are built from a potential, the linear confining potential leads to the linear long range part of the trajectory, which corresponds to the spectrum of the excited states of the mesons, and the one gluon exchange short range potential leads to the saturating part of the trajectory~\cite{Ser94}.  
In that sense, the exchange of the linear part of the trajectory can be viewed as the exchange of two quarks strongly interacting in a non perturbative way, while the exchange of the saturating part can be viewed as the  exchange of two quarks  interacting by the exchange of a single perturbative hard gluon. Saturating trajectories provide us with an economic and effective way to incorporate hard scattering mechanisms~\cite{Co84,Gui97}. Since in this channel we are left with no free parameters, the final analysis of the recent JLab data will put stringent constraints on this conjecture.

\begin{figure}[h]
\centerline{\epsfig{file=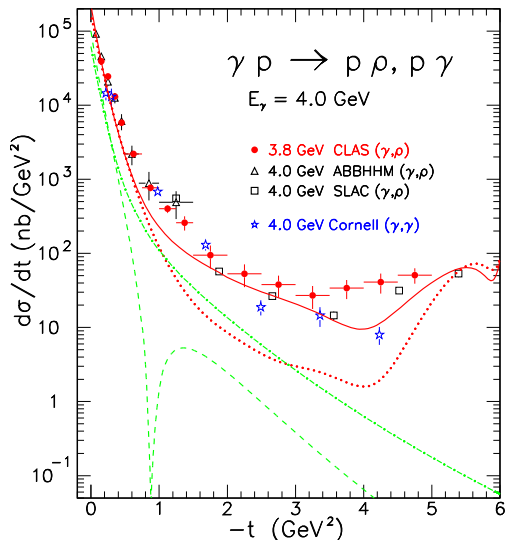,height=2.8in}}
\caption{The differential cross section of the $p(\gamma, \rho)p$
reaction in the JLab energy range.  Dotted (full) curve: full model
with linear (saturating) Regge trajectories. Dot-dashed (dashed)
curves: two gluon exchange with (without) correlations. Data for Compton
scattering has been renormalized by $f_V^2/4\pi \alpha_{em}$.}
\label{rho_slac}
\end{figure}

Fig.~\ref{rho_slac} shows the latest data~\cite{Ba01} obtained at JLab
in the $\rho$ photoproduction channel. At low $-t$ the good agreement
with the data is obtained adding, on top of the two gluon exchange
amplitude, the exchange of the $f_2$ and $\sigma$ Regge trajectories,
while the rise at high $-t$ comes from the $u$-channel exchange of the
$N$  and $\Delta$ Regge trajectories (see ref.~\cite{La00}). At
intermediate $-t$, the two gluon exchange contribution misses the data
by about one order of magnitude. The agreement is restored when a saturating trajectory ($\alpha_{\sigma}^{sat}(t)= \alpha_{\pi}^{sat}(t-0.25))$ is used instead of a linear trajectory, for the $\sigma$ meson. Since the $f_2$ meson Regge trajectory is non degenerated~\cite{La00}, its contribution vanishes when $\alpha_{f_2}=-1$: saturating it does not help. Due to the value of the coupling constants, pion exchange does not contribute to the $\rho$ channel. 

Such a model is consistent with the $s^{-7}$ asymptotic behavior of the $\theta
=90^{\circ}$ cross section of the $p(\gamma, \rho)p$ reaction depicted
in Fig.~\ref{rho_90}. Since the bulk of the cross section comes from
the exchange of the $\sigma$ Regge saturated trajectory, the two gluon
exchange has little effect on the final result, except at large $s$
where the onset of a definite power--like behavior may be delayed by a
gaussian propagator.

Finally, the same mechanism leads also to a good accounting  (Fig.~\ref{rho_slac}) of the angular distribution of Compton scattering at $E_{\gamma}=4$ GeV. A more comprehensive set of data will soon come from JLab, up to $E_{\gamma}=5.7$~GeV.

A consistent picture is emerging from the study of exclusive photoproduction of vector meson and Compton scattering. At low momentum transfer, it relies on diffractive scattering of the hadronic contents of the photon, in a wide energy range from threshold up to the HERA energy domain. At higher momentum transfer, it relies on a partonic description of hard scattering mechanisms which provides us with a bridge with Lattice Gauge calculations. The dressed gluon and quark propagators have already  been estimated on lattice. One may expect that correlated constituent quark wave functions (at least their first moments) will soon be available from lattice. An estimate, on lattice, of the saturating part of the Regge trajectories is definitely called for. 

To day, JLab is the only laboratory which allows to explore this regime, thanks to its high luminosity.
Its current operation, at 4--6 GeV, has already revealed a few
jewels. Its energy upgrade to 12 GeV will permit, among others, a more
comprehensive study of the onset of asymptotic hard scattering regime
and of the role of  correlations between quarks. 

This work is partially (FC) funded by European Commission IHP program (contract HPRN-CT-2000-00130).


\begin{thebibliography}{9}

\bibitem{La01} J.-M. Laget, {\it Nucl. Phys.} {\bf A} (in press); 
hep-ph/0107208.

\bibitem{La00} J.-M. Laget, {\it Phys. Lett.} {\bf B489} (2000) 313.
 
\bibitem{Va97} M. Vanderhaeghen, P. Guichon and J. Van de Wiele, 
{\it Nucl. Phys.} {\bf A622} (1997) 144c.

\bibitem{Di99} M. Diehl {\it et al.,Eur. Phys. J.} {\bf C8} (1999) 409.

\bibitem{Ra98} A.V. Radyushkin, {\it Phys. Rev.} {\bf D58} (1998) 114008.

\bibitem{Hu00} H. W. Huang and  Kroll, {\it Eur. Phys. J.} 
{\bf C17} (2000) 423.

\bibitem{LANDSHOFF87} P.V. Landshoff and O. Nachtmann, 
Z. Phys. {\bf C35} (1987) 405.

\bibitem{CUDELL94} J.R. Cudell and B.U. Nguyen, 
Nucl. Phys. {\bf B420} (1994) 669.

\bibitem{An00} E. Anciant {\it et al., Phys. Rev. Lett.} {\bf 85} (2000) 4682.

\bibitem{Br00} J. Breitweig {\it et al., Eur. Phys. J.} {\bf C14} (2000) 213.

\bibitem{BOLZ96} J. Bolz and P. Kroll, Z. Phys. {\bf A356} (1996) 327.

\bibitem{ZHITNITSKY98} A. Zhitnitsky, hep-ph/9801228.

\bibitem{CANO01} F. Cano and J.-M. Laget, {\it Nucl. Phys.} {\bf A}
(in press);  hep-ph/0107091.

\bibitem{WILLIAMS01} A.G. Williams {\it et al.},  hep-ph/0107029. 

\bibitem{LAGET95} J.-M. Laget and R. Mendez--Galain, Nucl. Phys. {\bf
A581} (1995) 397.

\bibitem{CORNWALL82} J.M. Cornwall, Phys. Rev. {\bf D26} (1982) 1453.

\bibitem{LEINWEBER99} D.B. Leinweber {\it et al.},  {\it Phys. Rev.}
{\bf D60} (1999) 094507.

\bibitem{Ja89} R. L. Jaffe, {\it Phys. Lett.} {\bf B229} (1989) 275.

\bibitem{Gui97} M. Guidal, J.-M. Laget and M. Vanderhaeghen, 
{\it Nucl. Phys.} {\bf A627} (1997) 645.

\bibitem{Ser94} M.N. Sergeenko, {\it Z. Phys.} {\bf C64} (1994) 315.

\bibitem{Co84} P.D.B. Collins and P.J. Kearney, {\it Z. Phys.} {\bf C22} (1984) 277.

\bibitem{Ba01} M. Battaglieri {\it et al., Phys. Rev. Lett.} {\bf 87}
(2001) 172002.

\end{thebibliography}
\end{document}